\documentclass[11pt]{article}
\usepackage{amsmath,amsxtra}
\usepackage{amscd}
\usepackage[dvips]{graphicx}
\usepackage{graphicx}
\usepackage{latexsym}
\usepackage{latexsym}
\usepackage{amsmath}
\date{\nonumber}
\begin{document}
\date{}
%  Theorems, Lemmas and the like, should be typeset in italic
\newtheorem{theorem}{Theorem}
\newtheorem{proposition}{Proposition}
\newtheorem{lemma}{Lemma}
\newtheorem{definition}{Definition}

\renewcommand{\theequation}
{\arabic{section}.\arabic{equation}}
%%%%% END DOCUMENT SPECIFIC DEFINITIONS
%\renewcommand{\square}{\hfill$\Box$\vspace{2ex}}
%\renewcommand{\Theta}{\Ta}

\renewcommand{\theequation}
{\arabic{section}.\arabic{equation}}
\title{The Degasperis-Procesi equation with self-consistent sources}
\author{ Yehui Huang$^{1}$\footnote{Corresponding author: Yehui Huang, Tel: +39-3318082062, e-mail:huangyh@mails.tsinghua.edu.cn}, Yunbo
Zeng$^{1}$\footnote{yzeng@math.tsinghua.edu.cn} and Orlando
Ragnisco$^{2,}$$^{3}$\footnote{ragnisco@fis.uniroma3.it}
\\ $^{1}$ Department of Mathematical Sciences,\\ Tsinghua University, Beijing, 100084, P.R. China\\
$^{2}$ Department of Physics, Universit$\grave{a}$ Roma TRE, Rome,
00146, Italy\\$^{3}$ I.N.F.N., sezione of Roma TRE, Rome, 00146,
Italy\\} \maketitle
\begin{abstract}
\indent The Degasperis-Procesi equation with self-consistent
sources(DPESCS) is derived. The Lax representation of the DPESCS is
presented. The conservation laws for DPESCS are constructed. The
peakon solution of DPESCS is obtained by using the method of
variation of constants.
\end{abstract}


 \vskip .1cm
\section{Introduction}
\setcounter{equation}{0} Soliton equations with self-consistent
sources (SESCS) have attracted much attention in recent years. They
are important integrable models in many fields of physics, such as
hydrodynamics, state physics, plasma physics, etc
\cite{1}-\cite{18}. For example, the KdV equation with
self-consistent sources describes the interaction of long and short
capillary-gravity waves \cite{5}. The KP equation with
self-consistent sources describes the interaction of a long-wave
with a short wave packet propagating on the $x-y$ plane at some
angle to each other \cite{2}. The nonlinear Schr$\ddot{o}$dinger
equation with self-consistent sources represents the nonlinear
interaction of an electrostatic high-frequency wave with the ion
acoustic wave in a two component homogeneous plasma \cite{6}. The
constrained flows of soliton hierarchy may be regarded as the
stationary systems of the corresponding integrable hierarchy with
self-consistent sources \cite{15}-\cite{18}. Since the Lax
representation for constrained flows can be deduced from the adjoint
representation of the Lax representation of soliton equation
\cite{14}, there is a naturally way to find the zero-curvature
representation for SESCS \cite{15}-\cite{18}. From this observation,
the soliton equations with self-consistent sources may be viewed as
integrable generalizations of the original soliton equations. A
systematic way to construct the soliton equations with
self-consistent sources is proposed in \cite{15}-\cite{18}.

The Camassa-Holm equation, which was implicitly contained in the
class of multi-Hamiltonian systems introduced by Fuchssteiner and
 Fokas in \cite{20} and explicitly derived as a shallow water wave equation by Camassa and
Holm in \cite{21}, has the form
\begin{equation}
u_t+2wu_x-u_{xxt}+3uu_x=2u_xu_{xx}+uu_{xxx}.
\end{equation}
Since the works of Camassa and Holm, this equation has become a
well-known example of integrable systems and has been studied from
many different points of view.

It is a natural question to ask whether there are other third order
dispersive PDEs sharing the integrability properties of Camassa-Holm
equation. An answer has been given in \cite{19} where the method of
asymptotic integrability was applied to a family of third-order
dispersive PDE conservation laws
\begin{equation}
u_t+c_0u_x+\gamma
u_{xxx}-\alpha^2u_{xxt}=(c_1u^2+c_2u_x^2+c_3uu_{xx})_x,
\end{equation}
where $\alpha$, $c_0$, $c_1$, $c_2$, $c_3$ and $\gamma$ are some
arbitrary constants. Only three equations in this family satisfy the
asymptotic integrability conditions. They are the KdV equation, the
Camassa-Holm equation and the following new equation
\begin{equation}
u_t+u_x+6uu_x+u_{xxx}-\alpha^2(u_{xxt}+\frac{9}{2}u_xu_{xx}+\frac{3}{2}uu_{xxx})=0.
\end{equation}
By a coordinate transformation, this new equation could be written
as \cite{22}
\begin{equation}
u_t-u_{xxt}+4uu_x=3u_xu_{xx}+uu_{xxx}.
\end{equation}

This encourages us to study the family of equations \cite{22}
\begin{equation}
u_t-u_{xxt}+(b+1)uu_x=bu_xu_{xx}+uu_{xxx}.
\end{equation}
All the equations in this family have peakon solutions of the form
\begin{equation}
u=\sum_{j=1}^N{p_j(t)e^{-|x-q_j(t)|}},
\end{equation}
where $p_j$ and $q_j$ satisfy the dynamical system
\begin{equation}
\dot{p_j}=-(b-1)\frac{\partial G_N}{\partial
q_j},\quad\dot{q_j}=\frac{\partial G_N}{\partial p_j},
\end{equation}
the generating function $G_N$ reading
\begin{equation}
G_N=\frac{1}{2}\sum_{j,k=1}^N{p_jp_ke^{-|q_j-q_k|}}.
\end{equation}

Let $m=u-u_{xx}$. The equation (1.4) could be written as
\begin{equation}
m_t+m_xu+3mu_x=0,
\end{equation}
which is called the Degasperis-Procesi (DP) equation. It is showed
in \cite{23} that both the Camassa-Holm and the DP equation are
derived as members of a one-parameter family of asymptotic shallow
water approximations to the Euler equations. They describe the
unidirectional propagation of nonlinear shallow-water waves. The DP
equation has a third order Lax pair and a bi-Hamiltonian structure.
The existence of its global solutions are considered in \cite{24}. A
new integrable hierarchy was extended from the DP equation in
\cite{25}. In \cite{26}, the N-peakon solutions of the DP equation
have been derived. The N-soliton solutions of the DP equation are
obtained in \cite {27}. Interesting results on the solutions of this
equation have also been obtained in \cite{28}-\cite{32}.

The soliton equation with self-consistent sources was firstly
studied by Melnikov in \cite{1}-\cite{3}. The problem of finding
soliton solutions or other specific solutions for equations with
self-consistent sources has been considered in the past by several
authors \cite{4}-\cite{18}. The present paper falls in that line of
research, aiming at deriving Degasperis-Procesi equation with
self-consistent sources (DPESCS) and to find for its special
explicit solutions.

We first construct the high-order constrained flow of the DP
equation. Based on it we establish the DPESCS by regarding the
constrained flow of DP equation as the stationary equation of DPESCS
in the same way as in \cite{15}-\cite{18}. The Lax pair of the
DPESCS is obtained, which means that the DPESCS is Lax integrable.
In \cite{33}-\cite{34} we pointed out that soliton equations with
self-consistent sources can be regarded as soliton equations with
non-homogeneous terms, and accordingly proposed to look for explicit
solutions by using the methods of variation of constants. Applying
this technique to DPESCS we have been able to find its peakon
solutions and peakon-antipeakon solution.

This paper is organized as follows. In section 2, we extend DP
equation including self-consistent sources and construct its Lax
representation. In section 3 we derive its conservation laws. In
section 4, the peakon and the peakon-antipeakon solution are
obtained. In section 5, we mention some open problems.

\section{The DPESCS and its Lax pair}
\subsection{The DPESCS}
\setcounter{equation}{0} First we construct the high-order
constrained flows of the DP equation, then establish the DPESCS and
describe how to derive the Lax representation for the DPESCS.

It is known that the Lax pair for DP equation (1.9) is \cite{22}
\begin{subequations}
\begin{eqnarray}
\psi_{xxx}&=&\psi_x-m\lambda\psi,\\
\psi_t&=&-\frac{1}{\lambda}\psi_{xx}-u\psi_x+(u_x+\frac{2}{3\lambda})\psi.
\end{eqnarray}
\end{subequations}

Consider the following equations obtained from the spectral problem
and its formal adjoint problem for $n$ distinct real $\lambda_j$.
\begin{subequations}
\begin{eqnarray}
q_{j,xxx}&=&q_{j,x}-m\lambda_jq_j,\quad j=1,\cdots,n,\\
r_{j,xxx}&=&r_{j,x}+m\lambda_jr_j,\quad j=1,\cdots,n.
\end{eqnarray}
\end{subequations}

It is not difficult to find that
\begin{equation}
\frac{\delta\lambda_j}{\delta m}=-\lambda_jq_jr_j,\quad
 j=1,\cdots,n.
\end{equation}

It is known that the DP equation possesses a bi-hamiltonian
structure \cite{22}, namely:
\begin{equation}\label{eq:CHH}
m_t=B_1\frac{\delta H_1}{\delta m}=B_0\frac{\delta H_0}{\delta m},
\end{equation}
where
\begin{eqnarray}
B_0&=&m^{2/3}\partial_xm^{1/3}(\partial_x-\partial_x^3)^{-1}m^{1/3}\partial_xm^{2/3},\\
B_1&=&\partial_x(1-\partial_x^2)(4-\partial_x^2),\\
H_0&=&-\frac{2}{9}\int{m dx},\\
H_1&=&-\frac{1}{6}\int{u^3dx}.
\end{eqnarray}

The high-order constrained flow of the DP equation is obtained from
(2.2) for $n$ distinct $\lambda_j$, requiring that the "potential"
$m$ obeys the following constraint
\begin{subequations}
\begin{eqnarray}
&&B_1(\frac{\delta H_1}{\delta
m}-\sum_{j=1}^n\alpha_j\frac{\delta\lambda_j}{\delta m})=0,\\
&&q_{j,xxx}=q_{j,x}-m\lambda_jq_j,\\
&&r_{j,xxx}=r_{j,x}+m\lambda_jr_j\quad j=1,\cdots,n,
\end{eqnarray}
\end{subequations}
where $\alpha_j$, $j=1,\ldots,n$ are arbitrary constants.

According to the approach proposed in \cite{15}-\cite{18}, the
DPESCS consists of the following equation
\begin{equation}
m_t=B_1(\frac{\delta H_1}{\delta
m}-\sum_{j=1}^n\alpha_j\frac{\delta\lambda_j}{\delta m})\notag
\end{equation}
and the equations (2.2), which by using (2.3) and taking
$\alpha_j=-\frac{1}{6}$ leads to the DPESCS
\begin{subequations}
\begin{eqnarray}\label{eq:CHESH}
m_t&=&-um_x-3u_xm-\frac{1}{6}\sum_{j=1}^n\partial(1-\partial^2)(4-\partial^2)(\lambda_jq_jr_j),\\
q_{j,xxx}&=&q_{j,x}-m\lambda_jq_j,\\
r_{j,xxx}&=&r_{j,x}+m\lambda_jr_j,\quad j=1,\cdots,n.
\end{eqnarray}
\end{subequations}

\subsection{The Lax representation of the DPESCS}
Comparing the DPESCS to the DP equation, we may assume that the Lax
presentation of the DPESCS has the form
\begin{subequations}
\begin{eqnarray}
\psi_{xxx}&=&\psi_x-m\lambda\psi,\\
\psi_t&=&-\frac{1}{\lambda}\psi_{xx}-u\psi_x+(u_x+\frac{2}{3\lambda})\psi+a\psi+b\psi_x+c\psi_{xx},
\end{eqnarray}
\end{subequations}
where $a$, $b$ and $c$ are some functions of $q_j$ and $r_j$ to be
determined. Requiring that under (2.10b) and (2.10c) the
compatibility condition of (2.11a) and (2.11b), namely
$\psi_{xxxt}=\psi_{txxx}$, leads to DPESCS (2.10a) enables us to
find that
\begin{subequations}
\begin{eqnarray}
&&a_x-a_{xxx}+3b_xm\lambda+bm_x\lambda+3c_{xx}m\lambda+3c_xm_x\lambda+cm_{xx}\lambda\notag\\
&&=\lambda\sum_{j=1}^{n}{\alpha_j\partial(1-\partial^2)(4-\partial^2)(q_jr_j)},\\
&&-3a_{xx}-b_{xxx}-2b_x-3c_{xx}+3c_xm\lambda+2cm_x\lambda=0,\\
&&3a_x+3b_{xx}+2c_x+c_{xxx}=0.
\end{eqnarray}
\end{subequations}
From (2.10b) and (2.10c) we obtain two identities
\begin{subequations}
\begin{eqnarray}
&&\partial(1-\partial^2)(4-\partial^2)(q_jr_j)=-3\lambda_j(m_xW(q_j,r_j)+3mW(q_j,r_j)_x),\\
&&W(q_j,r_j)_{xxx}-W(q_j,r_j)_x=\lambda_j(2m_xq_jr_j+3m(q_jr_j)_x),
\end{eqnarray}
\end{subequations}
where $W(q_j,r_j)=q_jr_{j,x}-q_{j,x}r_j$ is the usual Wronskian
determinant.

From (2.12c) we obtain that
\begin{equation}
3a+3b_x+2c+c_{xx}=0.
\end{equation}
Then (2.14) together with (2.12b) lead to
\begin{subequations}
\begin{eqnarray}
&&a_{xxx}-a_x=\frac{1}{6}\partial(1-\partial^2)(4-\partial^2)c+(cm_x\lambda+\frac{3}{2}c_xm\lambda)_x,\\
&&2((b+\frac{1}{2}c_x)_x-(b+\frac{1}{2}c_x)_{xxx})=2cm_x\lambda+3c_xm\lambda.
\end{eqnarray}
\end{subequations}
With the relations above we can rewrite (2.12a) as
\begin{eqnarray}
&&-\frac{1}{6}\partial(1-\partial^2)(4-\partial^2)c+\lambda(3m(b+\frac{1}{2}c_x)_x+m_x(b+\frac{1}{2}c_x))\notag\\
&&=\lambda\sum_{j=1}^{n}{\alpha_j\partial(1-\partial^2)(4-\partial^2)(q_jr_j)}.
\end{eqnarray}
Here we suggest that
\begin{subequations}
\begin{eqnarray}
&&c=\sum_{j=1}^n{A_jq_jr_j},\\
&&b+\frac{1}{2}c_x=\sum_{j=1}^n{B_jW(q_j,r_j)},
\end{eqnarray}
\end{subequations}
where $A_j$, $B_j$, $j=1,\ldots,n$ are some undetermined constants.

Finally with some calculations to determine $A_j$ and $B_j$ we solve
(2.12) for $a$, $b$, $c$ yielding:
\begin{subequations}
\begin{eqnarray}
a&=&\sum_{j=1}^n{\frac{1}{6}\frac{\lambda\lambda_j^2}{\lambda_j^2-\lambda^2}(3\lambda(q_jr_{j,xx}-q_{j,xx}r_j)-4\lambda_jq_jr_j-2\lambda_j(q_jr_j)_{xx})},\\
b&=&\sum_{j=1}^n{-\frac{1}{2}\frac{\lambda\lambda_j^2}{\lambda_j^2-\lambda^2}(\lambda(q_jr_{j,x}-q_{j,x}r_j)+\lambda_j(q_jr_j)_x)},\\
c&=&\sum_{j=1}^n{\frac{\lambda\lambda_j^3}{\lambda_j^2-\lambda^2}q_jr_j}.
\end{eqnarray}
\end{subequations}
In this way, we obtain the Lax pair for (2.10a) under(2.10b) and
(2.10c) as follows
\begin{subequations}
\begin{eqnarray}\label{eq:CHESLAX}
\psi_{xxx}&=&\psi_x-m\lambda\psi,\\
\psi_t&=&-\frac{1}{\lambda}\psi_{xx}-u\psi_x+(u_x+\frac{2}{3\lambda})\psi\notag\\
&&+(\sum_{j=1}^n{\frac{1}{6}\frac{\lambda\lambda_j^2}{\lambda_j^2-\lambda^2}(3\lambda(q_jr_{j,xx}-q_{j,xx}r_j)-4\lambda_jq_jr_j-2\lambda_j(q_jr_j)_{xx})})\psi\notag\\
&&+(\sum_{j=1}^n{-\frac{1}{2}\frac{\lambda\lambda_j^2}{\lambda_j^2-\lambda^2}(\lambda(q_jr_{j,x}-q_{j,x}r_j)+\lambda_j(q_jr_j)_x)})\psi_x\notag\\
&&+(\sum_{j=1}^n{\frac{\lambda\lambda_j^3}{\lambda_j^2-\lambda^2}q_jr_j})\psi_{xx},
\end{eqnarray}
\end{subequations}
which means that the DPESCS is Lax integrable.

\section{The infinite set of conservation laws for the DPESCS}
\setcounter{equation}{0}With the help of the Lax representation of
the DPESCS, we could find the conservation laws for the DPESCS by a
well-known method. First we assume that $m$, $u$ and its derivatives
tend to 0 when $|x|\rightarrow\infty$, and assume that $q_j$, $r_j$
and its derivatives tends to 0 when $x\rightarrow-\infty$. Set
\begin{equation}
\Gamma=\frac{\psi_x}{\psi},
\end{equation}
then the identity
\begin{equation}
\frac{\partial}{\partial t}(\frac{\partial\ln\psi}{\partial
x})=\frac{\partial}{\partial x}(\frac{\partial\ln\psi}{\partial
t})\notag
\end{equation}
together with (2.11) implies that DPESCS has the following
conservation law:
\begin{equation}
\frac{\partial}{\partial t}(\Gamma)=\frac{\partial}{\partial
x}(\frac{\psi_t}{\psi})=\frac{\partial}{\partial
x}(u_x+a-(u+b)\Gamma-(\frac{1}{\lambda}+c)(\Gamma_x+\Gamma^2)),
\end{equation}
where $a$, $b$ and $c$ are given by (2.14). Here we define that
$\Omega=u_x+a-(u+b)\Gamma-(\frac{1}{\lambda}+c)(\Gamma_x+\Gamma^2)$.
Using (2.11a) gives rise to
\begin{equation}
\Gamma-\Gamma_{xx}=m\lambda+3\Gamma\Gamma_x+\Gamma^3.
\end{equation}

We can have two kinds of expansions for $\Gamma$ in the power series
of $\lambda$. The first expansion is in positive powers of $\lambda$
\cite{22},
\begin{subequations}
\begin{eqnarray}
\Gamma&=&\sum_{k=0}^\infty h_k\lambda^{k+1},\\
\Omega&=&\sum_{k=0}^\infty g_k\lambda^{k+1}.
\end{eqnarray}
\end{subequations}
We note that the odd densities $h_{2k+1}$ are exact derivatives. The
first two densities are $h_0=u$, $h_2=u^3$, which yield the
conserved quantities $H_0$ and $H_1$ in (2.7) and (2.8). The
following densities are nonlocal because that would be an inverse of
the operator $(1-\partial_x^2)$ in the sequence.

The second expansion could be
\begin{subequations}
\begin{eqnarray}
\Gamma&=&\sum_{k=0}^\infty\mu_k\lambda^{\frac{1-k}{3}},\\
\Omega&=&\sum_{k=0}^\infty \omega_k\lambda^{\frac{1-k}{3}}.
\end{eqnarray}
\end{subequations}

With the relations of (3.2) and (3.3), we can obtain the infinite
densities and fluxes of the conservation laws. For brevity, we omit
the recursion relations here.

After some calculations, we can find that $\mu_k$ with odd
subscripts are derivatives of some functions. So we define
$H_{-s}=\int{\mu_{2s-2}}dx$, which are taken as the conserved
quantities. We could find the first few conserved quantities given
by $\mu_0$, $\mu_2$ are as follows
\begin{subequations}
\begin{eqnarray}
H_{-1}&=&\int m^{1/3} dx,\\
H_{-2}&=&\frac{1}{27}\int(m_x^2m^{-7/3}+9m^{-1/3})dx.
\end{eqnarray}
\end{subequations}

The corresponding flux of the conservation laws are
\begin{subequations}
\begin{eqnarray}
G_{-1}&=&m^{1/3}(-u+\frac{1}{2}\sum_{j=1}^n\lambda_j^2(q_jr_{j,x}-q_{j,x}r_j)),\\
G_{-2}&=&\frac{1}{2}u_x\sum_{j=1}^n\lambda_j^2(q_jr_{j,xx}-q_{j,xx}r_j))-\frac{1}{3}m_xm^{-1}(-u+\frac{1}{2}\sum_{j=1}^n\lambda_j^2(q_jr_{j,x}-q_{j,x}r_j)).
\end{eqnarray}
\end{subequations}

As the space part of the Lax pair is the same as that of DP
equation, the densities of the conservation laws are also the same.
Of course, as the time part is different, the fluxes will be also
different.

\section{Solution of the DPESCS}
\subsection{One peakon solution of the DPESCS}
\setcounter{equation}{0}As mentioned in the introduction, we will
construct peakon solutions for DPESCS by the method of variation of
constants.

The DP equation has peakon solutions \cite{22}. The one peakon is
\begin{equation}\label{eq:peak}
u=ce^{-|x-ct+\alpha|},
\end{equation}
where $\alpha$ is an arbitrary constant. The corresponding
eigenfunction of (4.1) is
\begin{subequations}
\begin{eqnarray}
q&=&\beta[sgn(x-ct+\alpha)(e^{-|x-ct+\alpha|}-1)-1],\\
r&=&\beta[sgn(x-ct+\alpha)(e^{-|x-ct+\alpha|}-1)+1],
\end{eqnarray}
\end{subequations}
where $\beta$ is arbitrary constant as well.

Taking $\alpha$ and $\beta$ in (4.1) and (4.2) to be time-dependent
$\alpha(t)$ and $\beta(t)$ and requiring that
\begin{subequations}
\begin{eqnarray}
u&=&ce^{-|x-ct+\alpha(t)|},\\
q&=&\beta(t)[sgn(x-ct+\alpha(t))(e^{-|x-ct+\alpha(t)|}-1)-1],\\
r&=&\beta(t)[sgn(x-ct+\alpha(t))(e^{-|x-ct+\alpha(t)|}-1)+1]
\end{eqnarray}
\end{subequations}
satisfies the DPESCS (2.10) for $n=1$. We find that
$c=\frac{1}{\lambda_1}$, $\alpha(t)$ can be an arbitrary function of
$t$ and $\beta(t)=\sqrt{\alpha'(t)c}$. So we have the one peakon
solution for (2.9) with $n=1$, $\lambda_1=\lambda=\frac{1}{c}$
\begin{subequations}
\begin{eqnarray}
u&=&ce^{-|x-ct+\alpha(t)|},\\
q&=&c\sqrt{\alpha'(t)}[sgn(x-ct+\alpha(t))(e^{-|x-ct+\alpha(t)|}-1)-1],\\
r&=&c\sqrt{\alpha'(t)}[sgn(x-ct+\alpha(t))(e^{-|x-ct+\alpha(t)|}-1)+1].
\end{eqnarray}
\end{subequations}

The one peakon of the DPESCS also has a cusp at its peak, located at
$x=ct-\alpha(t)$. We note that, while for the one peakon solution of
the DP equation travels with speed $c$ and has a cusp at its peak of
height $c$, for the DPESCS, the cusp is still at its peak of height
$c$, but the speed of the wave is no longer a constant.

\subsection{The peakon-antipeakon solution of the DPESCS}
The DP equation has peakon-antipeakon solutions \cite{22}
\begin{equation}
u=\frac{-sgn(t)c}{1-e^{-2c|t|}}(e^{-|x+c|t|+\alpha|}-e^{-|x-c|t|+\alpha|}),
\end{equation}
where $\alpha$ is an arbitrary constant. The corresponding
eigenfunctions of (4.5) is
\begin{subequations}
\begin{eqnarray}
q&=&\beta\sqrt{\frac{c}{(1-e^{-2c|t|})}}(sgn(x+c|t|+\alpha)(e^{-|x+c|t|+\alpha|}-1)\notag\\
&&-sgn(x-c|t|+\alpha)(e^{-|x-c|t|+\alpha|}-1)),\\
r&=&\beta\sqrt{\frac{c}{(1-e^{-2c|t|})}}(sgn(x+c|t|+\alpha)(e^{-|x+c|t|+\alpha|}-1)\notag\\
&&+sgn(x-c|t|+\alpha)(e^{-|x-c|t|+\alpha|}-1)).
\end{eqnarray}
\end{subequations}

Taking $\alpha$ and $\beta$ in (4.5) and (4.6) to be time-dependent
$\alpha(t)$ and $\beta(t)$, and with the method of the variation of
constants, the peakon-antipeakon solution of the DPESCS with $n=1$
and $\lambda_1=\frac{1}{c}$ is
\begin{subequations}
\begin{eqnarray}
u&=&\frac{c}{1-e^{-2c|t|}}(e^{-|x+c|t|+\alpha(t)|}-e^{-|x-c|t|+\alpha(t)|}),\\
q&=&c\sqrt{\frac{-sgn(t)\alpha'(t)}{(1-e^{-2c|t|})}}(sgn(x+c|t|+\alpha(t))(e^{-|x+c|t|+\alpha(t)|}-1)\notag\\
&&-sgn(x-c|t|+\alpha(t))(e^{-|x-c|t|+\alpha(t)|}-1)),\\
r&=&c\sqrt{\frac{-sgn(t)\alpha'(t)}{(1-e^{-2c|t|})}}(sgn(x+c|t|+\alpha(t))(e^{-|x+c|t|+\alpha(t)|}-1)\notag\\
&&+sgn(x-c|t|+\alpha(t))(e^{-|x-c|t|+\alpha(t)|}-1)),
\end{eqnarray}
\end{subequations}
where $\alpha(t)$ is an arbitrary function of $t$.

\section{Conclusion}
As a concluding remark, we want to stress that the construction of N
peakon solutions for the DPESCS is still an open problem, as well as
for the case of the Camassa-Holm equation with self-consistent
sources. In fact, in the case of the Camassa-Holm equation and DP
equation a reciprocal transformation is used, but so far we have not
been able to extend it in the case of those equations with
self-consistent sources.

\section*{Acknowledgements}
This work was supported by the National Basic Research Program of
China (973 program) (2007CB814800) and the National Science
Foundation of China (Grant no 10601028). Yehui Huang would like to
thank "Antenna Lazio" for having a one-year fellowship to stay at
the Physics department of Roma Tre University.

\end{document}